\documentclass[structabstract]{an}
\usepackage{graphicx}
\usepackage{times}
\usepackage{amssymb} 
\def\nh{{N$_{\rm H}$}}

\begin{document}

\Pagespan{0}{}
\Yearpublication{??}%
\Yearsubmission{2012}%
\Month{?}%
\Volume{?}%
\Issue{?}%

\title{X--ray luminosity functions of different morphological and X--ray type AGN populations}

\author{M. Povi\'c\inst{1, 2},\fnmsep\thanks{Corresponding author:
  \email{mpovic@iaa.es}\newline}
A. M. P\'erez Garc\'ia\inst{1, 3},
M. S\'anchez-Portal\inst{5, 3},
A. Bongiovanni\inst{1, 4, 3},
J. Cepa\inst{4, 1},
M. Fern\'andez Lorenzo\inst{1, 2},
M. A. Lara-L\'opez\inst{1, 6},
J. Gallego\inst{7}, 
A. Ederoclite\inst{1, 8},
I. M\'arquez\inst{2},
J. Masegosa\inst{2},
E. Alfaro\inst{2}, 
H. Casta\~neda\inst{9}, 
J. I. Gonz\'alez-Serrano\inst{10}, 
\and J. J. Gonz\'alez\inst{11}}
\titlerunning{\textbf{X--ray luminosity functions of different AGN populations}}
\authorrunning{Povi\'c et al.}
\institute{
Instituto de Astrof\'isica de Canarias (IAC), La Laguna, Tenerife, Spain
\and 
Instituto de Astrof\'isica de Andaluc\'ia (IAA-CSIC), Granada, Spain
\and 
Asociaci\'on ASPID, Apartado de Correos 412, La Laguna, Tenerife, Spain
\and 
Departamento de Astrof\'isica, Universidad de La Laguna (ULL), La Laguna, Spain
\and 
Herschel Science Centre, ESAC/INSA, Madrid, Spain
\and 
Australian Astronomical Observatory (AAO), Sidney, Australia
\and 
Departamento de Astrof\'isica y CC. de la Atm\'osfera, Universidad Complutense de Madrid, Madrid, Spain
\and
Centro de Estudios de Física del Cosmos de Arag\'on (CEFCA), Teruel, Spain
\and 
Escuela Superior de F\'isica y Matem\'aticas, Intituto Polit\'ecnico Nacional (IPN), Mexico D.F, Mexico
\and 
Instituto de F\'isica de Cantabria, CSIC-Universidad de Cantabria, Santander, Spain
\and
Instituto de Astronom\'ia UNAM, M\'exico D.F, M\'exico
}
\received{?}
\accepted{?}
\publonline{later}


\abstract{%
Luminosity functions are one of the most important observational clues when studying galaxy evolution over cosmic time. In this paper we present the X--ray luminosity functions for X--ray detected AGN in the SXDS and GWS fields. The limiting fluxes of our samples are 9.0\,$\times$\,10$^{-15}$ and 4.8\,$\times$\,10$^{-16}$\,erg\,cm$^{-2}$\,s$^{-1}$ in the 0.5\,-\,7.0\,keV band in the two fields, respectively. We carried out analysis in three X--ray bands and in two redshift intervals up to z\,$\le$\,1.4. Moreover, we derive the luminosity functions for different optical morphologies and X--ray types. We confirm strong luminosity evolution in all three bands, finding the most luminous objects at higher redshift. However, no signs of density evolution are found in any tested X--ray band. We obtain similar results for compact and early-type objects. Finally, we observe the `Steffen effect', where X--ray type-1 sources are more numerous at higher luminosities in comparison with type-2 sources.}

\maketitle

\section{Introduction}
\label{sec_intro}

\indent \indent The luminosity function (LF) is among the fundamental tools for studying the formation and evolution of 
galaxies (e.g. Schechter 1976). Galaxies with an Active Galactic Nucleus (AGN) in their centres are 
among the most luminous sources in the Universe. Therefore, they provide unique information about the 
properties of high-redshift galaxies that is important for understanding how galaxies formed and evolved. However, 
the manner in which a galaxy enters the active phase, the duration of the activity, and the origin of 
the accreting material have not been clarified yet. The use of LFs to determine the 
distribution and evolution of the AGN accretion history has been shown to be essential for constraining 
models of formation and evolution of super-massive black holes (SMBHs), AGN feedback, fuelling me-\\chanisms, and 
the relation between active and non-active galaxies (e.g., Aird et al. 2010, La Franca et al. 2010, Fiore 
et al. 2012 and references therein). LF computation requires the availability of data for large samples of sources, covering a wide range of redshift and luminosity, in order to accurately determine the shape of the 
LF and its evolution over cosmic time.

\indent X--ray emission is one of the principal characteristics of AGN activity. Since X--ray data are relatively 
unaffected by absorption, they have been shown to be very useful for selecting large AGN samples. Hard and 
very-hard X--ray surveys are especially important, since they are able to detect significant populations of sources 
that are obscured in optical and NIR bands. Therefore, X--ray data provide the most complete and unbiased samples of 
AGN, and the X--ray luminosity function (XLF) can provide important knowledge about AGN evolution over cosmic time 
(e.g., Barger et al. 2005; Barger \& Hasinger et al. 2005).

\indent The AGN LF has been widely studied over past years, providing important constraints on 
models of galaxy formation and evolution. In the last decade, deep X--ray (e.g., Ueda et al. 2003; La Franca et al. 2005; Hopkins, Richards, \& Hernquist 2007; Silverman et al. 2008a, 2008b; Ebrero et al. 2009; Yencho et al. 2009; Aird et al. 2010 and references therein), optical (e.g., Boyle et al. 2000; Fan et al. 2001, 2004; Croom et al. 2004; Richards et al. 2005, 2006; Jiang et al. 2006; Cool et al. 2012), IR (Dai et al. 2009 and references therein; Wu et al. 2011 and references therein) and radio (e.g., Kaiser \& Best 2007; Kellermann et al. 2012 and references therein) surveys studied large samples of AGN over a broad range of redshift (z\,$\lesssim$\,6). These multiwavelength studies have provided a wider view of AGN properties. As already mentioned, X--ray selection provides one of the most complete samples of active galaxies; however they may miss the most obscured (Compton thick) sources. These sources can be observed in the  IR. In contrast, optical data are important for providing the redshift and AGN morphology. Current observations show that the \\AGN space density reaches a broad peak at z\,$\approx$\,1\,-\,3 (e.g. Hasinger et al. 2005; Silverman et al. 2008b; Aird et al. 2008; Brusa et al. 2009) followed by a decline of $\approx$\,2\,dex to the present day 
(e.g., Ueda et al. 2003; Hasinger et al. 2005; Barger et al. 2005). Moreover, X--ray studies (e.g., Page et al. 1997; Miyaji et al. 2000, 2001; La Franca et al. 2002; Cowie et al.
2003; Ueda et al. 2003; Hasinger et al. 2005; Aird et al. 2010 and references therein), but also optical, IR, and radio measurements (e.g., Hunt et al. 2004; Cirasuolo, Magliocchetti, \& Celotti 2005; Matute et al. 2006) indicate that the space density of low-luminosity AGN peaks at lower redshift than the bright AGN. In order to explain this effect, called 'cosmic downsizing' (Cowie et al. 2003; Merloni 2004; Heckman et al. 2004, Babi\'c et al. 2007; Fanidakis et al. 2012), most of the proposed models of AGN formation and evolution consider different forms of AGN feedback processes as the responsible mechanism (Di Matteo et al. 2005; Hopkins et al. 2006; Croton et al. 2006; Hopkins et al. 2008).

\indent Many studies of the AGN LF have been published, leading to different models of AGN evolution. However, their analyses were restricted to the evolution of the entire AGN population.In this work, 
in addition to a study of the entire AGN  sample, the X--ray luminosity 
function was also studied for different morphologies and X--ray 
obscurations for redshifts z\,$\le$\,1.4. The results presented here represent a pilot study of a much larger 
sample of AGN that will be carried out in the COSMOS field (Scoville et al. 2007) with the final aim 
of studying the evolution of different AGN triggering mechanisms over cosmic time. The paper is structured as follows: in Section~\ref{sec_obsdata}, we describe the data used in this work, including X--ray (Section~\ref{subsec_xray}), optical (Section~\ref{subsec_opt}), near- and mid-infrared (NIR and MIR) data (Section~\ref{subsec_ni_mir}), and provide brief summary of photometric redshift and k-correction estimations of the optical counterparts of X--ray emitters (Section~\ref{subsec_optcounter_z_kcorr}). In Section~\ref{sec_morph_xtype}, we describe the optical morphology and X--ray type classification of X--ray selected AGN in both fields. In Section~\ref{sec_LF}, we 
present the analysis of the XLFs, with the luminosity 
function measurements described in Section~\ref{sec_LF_measure} and discussion of the 
LFs of the whole AGN sample in Section~\ref{sec_LF_Xray_wholesample}, of the different 
morphological types in Section~\ref{subsec_LF_Xray_morph}, and of the different X--ray types in 
Section~\ref{subsec_LF_Xray_nucty}. Our study of the evolution of 
X--ray type-1 sour-ces is given in Section~\ref{subsec_LF_Xray_Xrayty1}. Finally, the results obtained in 
this work are summarized in Section~\ref{sec_summary}.

\indent The concordance cosmology with $\Omega_{\Lambda}$\,=\,0.7, $\Omega_{M}$\,=\,0.3, and H$_0$\,=\,70 km s$^{-1}$ Mpc$^{-1}$ is assumed. Unless otherwise specified, all magnitudes are given in the AB system (Oke \& Gunn 1983).

\section{Observational data}
\label{sec_obsdata}

\indent \indent This work involves sources  belonging to two fields: Su-baru/XMM-\textit{Newton} Deep Survey (SXDS) and Groth-West-phal Strip (GWS). These fields were selected for the OSI-RIS Tunable Emission Line Object (OTELO) survey (Cepa et al. 2008). This is a deep on-going survey of emission line sources that are being observed with tunable filters in the Optical System for Imaging and low Resolution Integrated Spectroscopy (OSIRIS) on the Gran Telescopio de Canarias (GTC\footnote{http://www.gtc.iac.es/}) telescope. Results presented in this paper complement studies of the AGN/QSO LFs carried out in the OTELO survey also using OSIRIS data. \\
\indent The SXDS is a large survey covering  $>$\,1\, square degree centred at RA\,=\,02$^h$18$^m$ and DEC = $-$05$^{\circ}$00' (Sekiguchi et al. in prep), and observed from X--ray to radio. \\ 
\indent The GWS field is a part of the Extended Groth Strip (EGS) centred at 
RA\,=\,14$^h$17$^m$ and DEC\,=\,+\,52$^{\circ}$30'. Broad band optical $BVRI$ observations were performed by the OTELO group (Cepa et al. 2008). The EGS was also observed by the All-Wavelength Extended Groth Strip International Survey (AEGIS\footnote{http://aegis.ucolick.org/}) team, with the aim of obtaining deep images and spectroscopic data from X--ray to radio, covering an area of 0.5\,-\,1\,square degree (Davis et al. 2007).\\
\indent We mainly use X--ray and optical data in this work. NIR and MIR data were used to complement the optical information for the estimation of photometric redshifts estimations (see Section~\ref{subsec_optcounter_z_kcorr}). The observational data are briefly described below.

\subsection{X--ray data}
\label{subsec_xray}

\indent \indent The SXDS field was observed with XMM-\textit{Newton} at seven pointings in the 0.2--10\,keV energy range (PI Michael G. Watson) and covering $\simeq$\,1.3 deg$^2$. The central observation is deepest with a nominal exposure time of $\sim$\,100\,ksec. It is surrounded by six shallower observations each with an exposure of $\sim$\,50\,ksec. \\
\indent The required European Photon Imaging Camera (EP-IC\footnote{http://xmm.esac.esa.int/external/xmm\_user\_support/documentation\\/technical/EPIC/}) data were collected 
from the XMM-\textit{Newton} v.5.0 scientific archive (XSA\footnote{http://xmm.esac.esa.int/xsa/}) and were processed 
by our team using standard procedures of the Science Analysis System \\(SAS\footnote{http://xmm.esac.esa.int/external/xmm\_data\_analaysis}) v7.1.2 
package, and the latest relevant Current Calibration Files (CCF). Source detection was based on the \texttt{edetect}
chain SAS procedure. In order to minimize spurious detections the likelihood threshold parameter\footnote{L = -ln(1 - P), where P is the probability to have a spurious detection due to random Poisson fluctuations.} was set at L\,=\,14. Thus the probability that the source is real exceeds  P\,=\,0.99 which implies less than one fake source per instrument, per pointing, and per band. A detailed description of X--ray data reduction and source detection is available in Povi\'c et al. (2012). We detected 1121 sources with fluxes greater than our flux detection limit of 9.0\,$\times$ 10$^{-15}$ erg cm$^{-2}$ s$^{-1}$ in the 0.5\,-\,7.0\,keV band. \\

\indent The GWS field was observed with \textit{Chandra} covering three continuous fields centred at the original HST GWS. All observations were carried out by the AEGIS collaboration using the Advanced CCD Imaging Spectrometer (AC-IS-I\footnote{http://cxc.harvard.edu/cal/Acis/}) instrument, with a total exposure time of $\sim$\,200 ksec in each field, covering an area of 0.08\,deg$^2$.\\
\indent All datasets were gathered from the \textit{Chandra} Data Arc-hive\footnote{http://asc.harvard.edu/cda/} and were processed by our team using the \textit{Chandra} Interactive Analysis of Observations (CIAO)\footnote{http://cxc.harvard.edu/ciao/} v4.0 package and Calibration Data Base (CALDB) v3.4.0. Standard reduction procedures were applied to produce new level 2 event files, restricted to the 0.5--8\,keV range in order to avo-id high background spectral regions. Good time intervals were defined by means of $\texttt{celldetect}$ and a threshold of S/N\,=\,3. Absolute astrometry was improved, where the average positional accuracy improvement is 6.3\% with respect to the original astrometry provided by the spacecraft attitude 
files, and the event files were merged. We applied the CIAO \texttt{wavdetect} Mexican-Hat wavelet source 
detection program to all bands. We set a significance threshold of  2\,$\times$\,10$^{-7}$, expecting about 12 spurious sources in the entire catalogue. The final catalogue has 639 unique X--ray emitters, with fluxes greater than the flux detection limit of 4.8\,$\times$\,10$^{-16}$ erg cm$^{-2}$ s$^{-1}$ in the 0.5\,-\,7.0\,keV band. A detailed description of X--ray data reduction and source detection can be found in Povi\'c et al. (2009).\\

\indent In this paper, we derive the XLF in the 0.5--2\,keV (soft) and 0.5--7\,keV (total) bands for both the SXDS and GWS samples, and in the 2-10\,keV  (hard) band for the SXDS sample. To separate between the X--ray unobscured and obscured 
sources (see Section~\ref{sec_morph_xtype}), we use HR$(2-4.5$keV/ $0.5-2$keV$)$ hardness ratio defined as:
\begin{equation} \label{eq_HR}
HR = \frac{CR(2-4.5\,keV) - CR(0.5-2\,keV)}{CR(2-4.5\,keV) + CR(0.5-2\,keV)},
\end{equation}
where CR is the count rate in a given energy band. \\

\begin{figure}[t]
\centering
\includegraphics[width=8.0cm,angle=0]{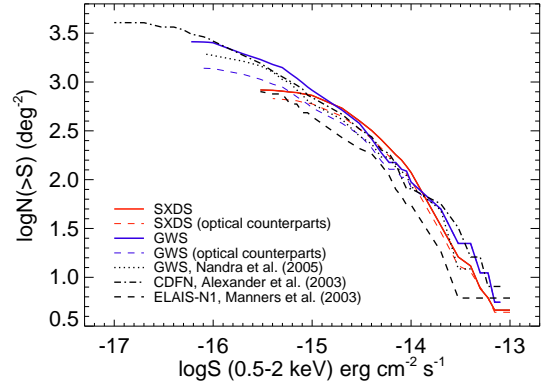}
\protect\caption[ ]{Cumulative logN-logS functions for the SXDS and GWS fields in the soft band, for all detected X--ray sources 
(red and blue thick solid lines, respectively) and for X--ray emitters with optical counterpart (red and blue thin dashed lines, respectively). Three distributions, with different exposure times and effective areas, are represented for comparison: GWS by Nandra et al. (2005) (black dotted line), CDFN (black dash-dot-dashed line), and ELAIS-N1 (black dashed line).
\label{fig_logn_logs}}
\end{figure}

\indent We tested the reliability of our X--ray number counts in the SXDS and GWS fields by comparing with number counts obtained in deeper (e.g. \textit{Chandra} Deep Field North (CDFN); Alexander et al. 2003) and less deep (e.g., ELAIS-N1; Manners et al. 2003) surveys. Figure~\ref{fig_logn_logs} shows the cumulative number count distributions (logN-logS) per square degree in the soft band. Two curves have been computed for the SXDS and GWS sample, one for the complete sample of X--ray detections and the other for the sample of X--ray detections with optical counterparts. For objects brighter than $\sim$\,4\,$\times$\,10$^{-14}$\,erg\,s$^{-1}$\,cm$^{-2}$ in the case of SXDS, and $\sim$\,10$^{-14}$\,erg\,s$^{-1}$\,cm$^{-2}$ in the case of GWS, both distributions (complete X--ray sample and X--ray sample with optical counterparts) coincide, while for fainter objects the density of X--ray sources with optical counterparts starts to decrease.  
In contrast, the difference in the density of X--ray sources between the distributions of all objects and optical counterparts is lower in the case of the SXDS sample, since the optical data are deeper in comparison with the GWS sample.
Since X--ray data are deeper in the GWS field, and since it is important to deal with the complete sample when measuring the LF, we will use the limiting fluxes corresponding to the completeness levels of SXDS samples in all analyses.

\subsection{Optical data}
\label{subsec_opt}

\indent \indent The SXDS optical imaging observations were carried out with the Suprime-Cam on the Subaru Telescope We observed
five continuous sub-fields  covering an area of 1.2 deg$^2$ with a total integration time of 133 hours (Furusawa et al. 2008). We downloaded and used publicly available catalogues from the SXDS web page\footnote{http://www.naoj.org/Science/SubaruProject/SXDS/index.html}, obtaining photometric information in five broad-band filters $B$, $V$, $R_c$, $i'$, and $z$, with limiting AB magnitudes of 28.4, 27.8, 27.7, 27.7, and 26.6, respectively. The total number of objects in each field is around 900000. \\ 

\indent GWS observations used in this work were carried out by the OTELO team using the Prime Focus Imaging Platform (PFIP) at the William-Herschel Telescope (WHT)\footnote{http://www.ing.iac.es/Astronomy/telescopes/wht/} covering a total area of 0.18\,deg$^2$. Three pointings were observed, each of them using the $B$, $V$, $R$, and $I$ broadband filters, and performing several exposures at each pointing of 600, 800, 900, or 1000\,sec depending on the filter. Standard reduction procedures have been applied. Resulting limiting AB magnitudes are 24.8, 24.9, 24.6 and 23.8 in $B$, $V$, $R$, and $I$ band, respectively. The total number of objects is around 44000. A detailed description of data reduction and optical source detection can be found in Cepa et al. (2008).

\subsection{NIR and MIR data}
\label{subsec_ni_mir}

\indent \indent NIR information for the SXDS field was taken from the UKIDSS DR2 catalogue\footnote{http://www.ukidss.org/archive/archive.html} (Hewett et al. 2006; Warren et al. 2007; Hodgkin et al. 2009).
This data has AB limiting magnitudes of 23.4, 23.4, and 22.9 in $J$, $H$, and $K$ bands, respectively, and covers an area of 8.75\,deg$^2$. MIR data for the SXDS field was obtained with the \textit{Spitzer} telescope as part of the XMM-LSS survey
which is  one of the areas surveyed within the \textit{Spitzer} Wide-area IR Extragalactic Legacy Survey (SWIRE) covering a total area of 9.1\,deg$^2$. We used the public Infrared Array Camera (IRAC\footnote{http://www.cfa.harvard.edu/irac/}) catalogue\footnote{http://swire.ipac.caltech.edu/swire/astronomers/data\_access.html} of approximately 250700 objects, with AB limiting magnitudes of 21.0, 21.1, 19.5, and 19.4 in the 3.6, 4.5, 5.8, and 8.0\,$\mu$m bands, respectively.

\indent We used publicly available NIR $K_s$ Palomar data\footnote{http://www.astro.caltech.edu/AEGIS/} for the GWS field. The catalogue contains $\sim$\,45000 objects wi-th an AB limiting magnitude of 20.6 covering an area of 0.7 deg$^2$. \textit{Spitzer}/IRAC EGS catalogue\footnote{http://www.cfa.harvard.edu/irac/egs/} is publicly available and has been used in this work to gather MIR data for GWS X--ray emitters. The raw catalogues contain (57.4, 48.0, 16.3, 13.6)\,$\times$\,10$^3$ objects, with AB limiting magnitudes of 23.2, 22.8, 21.3, and 21.1 in the four IRAC channels, respectively, and a total area of 0.33 square degrees.

\subsection{Redshifts and k-corrections of optical counterparts}
\label{subsec_optcounter_z_kcorr}

\indent \indent We cross-matched the optical data with X--ray catalogu-es using a search radius of 3$''$ and 2$''$ for SXDS and GWS fields, respectively. Applying the de Ruiter methodology (de Ruiter et al 1977; Povi\'c et al. 2009) we obtain a completeness of 99.9\% and 98.3\%, and a reliability of 76.2\% and 87.3\% in the SXDS and GWS fields, respectively. Due to the poor reliability in the SXDS field and, in order to decrease the number of spurious detections, we also applied the Sutherland \& Saunders (1992) methodology, measuring the reliability for all possible matches inside the selected radius, and finally selecting, as an optical counterpart, 
the object with the highest probability. This additional step drops the 
percent of possible fake identifications from 24\% to 6\%, in cases where 
multiple matches are detected with similar probabilities. We find 808 and 
340 X--ray sources with optical counterpart in the SXDS and GWS fields, respectively. The final covered areas are 1.2\,deg$^2$ and 0.08\,deg$^2$ in the SXDS and GWS fields, respectively. These are the areas that will be used for solid angle estimations in Section~\ref{sec_LF}, and cover the overlap area of all the data employed to identify X--ray sources and their optical counterparts, and to measure photometric redshifts. 

\begin{figure}[t]
\centering
\begin{minipage}[c]{.47\textwidth}
\includegraphics[width=6.0cm,angle=0]{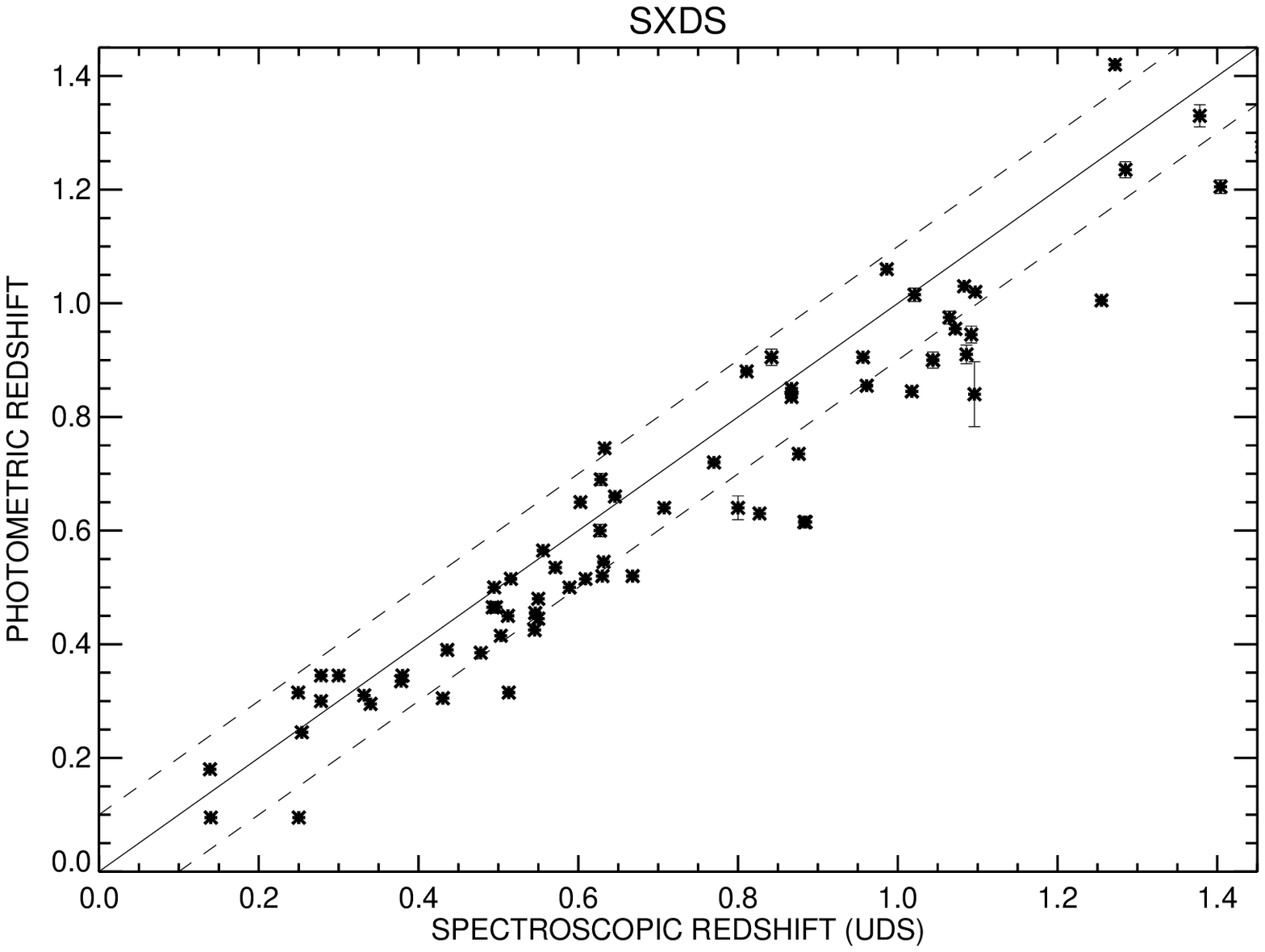}
\centering
\end{minipage}
\begin{minipage}[c]{.47\textwidth}
\includegraphics[width=6.0cm,angle=0]{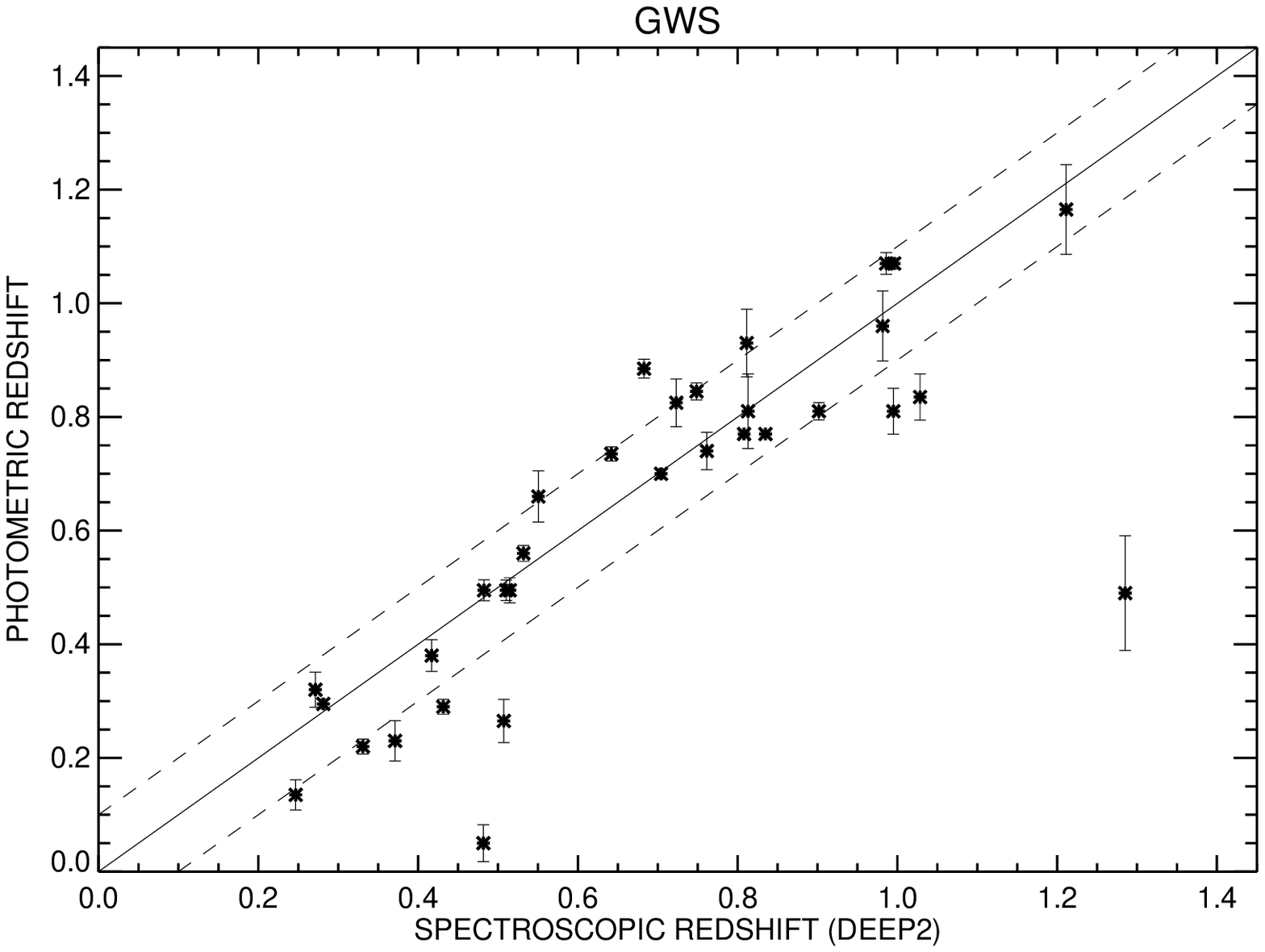}
\centering
\end{minipage}
\caption[ ]{Relation between our photometric and spectroscopic redshifts in the SXDS \textit{(top)} and GWS \textit{(bottom)} fields. Solid line shows the best correlation, when zphot=zspec, while the dashed lines show the $\pm$\,0.1 difference limits.
\label{fig_zphot_zspec}}
\end{figure}

\indent We used the \texttt{ZEBRA} (Feldmann et al. 2006) code to compute photometric redshifts for optical counterparts. Due to the small number of previously available spectroscopic redshifts in the SXDS field (fifteen objects), and the lack of NIR data in the GWS survey (only K data available), photometric redshifts were also derived using the \texttt{HyperZ} code (Bolzonella et al. 2000) as an additional check. We implemented and used the templates from the SWIRE\footnote{http://www.iasf-milano.inaf.it/$\sim$polletta/templates/swire\_templates.html} library (Polletta et al. 2007), including galaxies belonging to different morphological types, starbursts, different types of Seyfert galaxies, and type-1 and type-2 QSO templates. Comparing the results from \texttt{ZEBRA} and \texttt{HyperZ}, we accepted only sources where the redshift results from both codes agreed within less than 0.1 for z\,$<$\,1 and 0.2 for z\,$\geq$\,1. Following this criterion we derived photometric redshifts for 308 (38\%) and 203 (60\%) objects in the SXDS and GWS fields respectively. We focus our studies on the 208 (SXDS) and 169 (GWS) objects that have redshifts below 1.4 in order to derive an LF for sources with reliable redshift estimates. Additionally, we compared our photometric redshift in the SXDS field with the spectroscopic values recently provided by the UDS team\footnote{http://www.nottingham.ac.uk/astronomy/UDS/data/dr3.html}. The relation is presented in Figure~\ref{fig_zphot_zspec} (top panel). 73\% and 86\% of all objects have photometric--spectroscopic redshift differences below 0.1 and 0.15, respectively. In the GWS field the comparison between photometric and used spectroscopic values is presented in Figure~\ref{fig_zphot_zspec} (bottom panel). In this field, 66\% and 81\% of all objects show differences below 0.1 and 0.15, respectively. In the following analysis, we used the spectroscopic redshifts when available, and otherwise our photometric estimates (69\% and 83\% for samples from the SXDS and GWS fields, respectively).    

\indent The code \texttt{kcorrect} (Blanton et al. 2007) was used in both fields in order to obtain k-corrections for the optical fluxes. In contrast. X--ray fluxes were k-corrected by assuming a standard power law spectral energy distribution (SED) with $\Gamma$\,=\,1.8. No intrinsic absorption correction has been applied to the power law
SED for objects considered to be unobscured(see Section~\ref{sec_morph_xtype}) a fixed intrinsic absorption 
of \nh\,=\,1.0\,$\times$\,10$^{22}$ cm$^{-2}$ was applied (e.g., Silverman et al. 2005; Younes et al. 2012) for obscured sources.

\section{Morphological and X--ray classification}
\label{sec_morph_xtype}

\begin{figure}[t]
\centering
\includegraphics[width=8.3cm,angle=0]{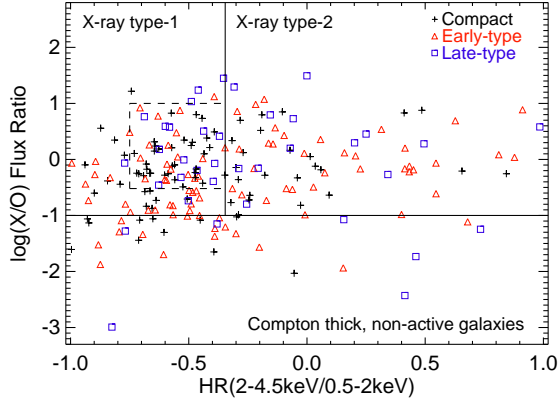}
\protect\caption[ ]{Relation between the X/O flux ratio and HR$(2-4.5$keV,$0.5-2$keV$)$, for X--ray emitters with optical counterparts in the SXDS and GWS fields. Sources belong to different morphological groups: compact (black crosses), early- (red triangles), and late-type (blue squares) galaxies. Solid lines separate X--ray type-1 (unobscured) and X--ray type-2 (obscured) regions and the area with logX/O\,$<$\,-1, where stars, non-active galaxies, and Compton thick AGN can be found (e.g. Fiore et al. 2003). The dashed line box presents the limits obtained by Della Ceca et al. (2004), presenting the `highest probability' region for finding X--ray type-1 AGN.
\label{im_logXO_HRh2s_sxds_gws_morph}}
\end{figure}

\indent \indent We used \texttt{SExtractor} (Bertin \& Arnouts 1996) and \texttt{galSVM} (Huertas-Company et al. 2008, 2009) codes to ev-aluate I-band morphologies for optical counterparts of X--ray emitters in the SXDS field. \texttt{SExtractor} was used for detection of compact sources (CLASS\_STAR\,$\ge$\,0.9), while we classified extended objects with \texttt{galSVM}. As an additional check we made a visual classification for the largest and brightest objects using the \texttt{IRAF/imexam} (Tody 1986, 1993) tool and isophotal contour diagrams. The 
final classifications include 22\% compact, 30\% early-type, 18\% late-type, 23\% poorly classified  and 7\% unidentified 
sources. Further details about morphological classifications in the SXDS field can be found in Povi\'c et al. (2012).\\
\indent We obtained morphologies for X--ray emitters with optical counterparts in the GWS field using \texttt{SExtractor} (Bertin \& Arnouts 1996; to select compact sources as in the SXDS field) and \texttt{GIM2D} (Simard et al. 1998) codes, and we performed again a visual inspection as an additional check. Morphology in the GWS field was determined prior to the \texttt{galSVM} code becoming publicly available. The final classifications were carried out in the R band (the I band was affected by fringing) and included: 42\% compact, 21\% early-type galaxies, 27\% late-type galaxies, and 10\% unidentified sources. The number of compact objects is significantly higher in this field probably due to the lower quality of optical data  compared to the SXDS field. Further details related with morphological classification in the GWS field can be found in Povi\'c et al. (2009). \\\\
\indent A coarse nuclear type classification was performed in both fields, based on a diagnostic diagram relating the X--ray-to-optical (X/O) flux ratio and the hardness ratio HR $(2-4.5$keV/$0.5-2$keV$)$. The X/O ratio was computed as ratio of fluxes $F_{\rm 0.5-4.5keV}/F_R$, where the optical flux $F_R$ was derived from the \texttt{SExtractor} auto magnitudes in the $R$ band. This flux ratio is useful to efficiently separate between active and non-active/Compton-thick galaxies (Alexander et al. 2001; Fiore et al. 2003; Civano et al. 2007), while the hardness ratio HR$(2-4.5$keV/$0.5-2$keV$)$, measured as in equation~(\ref{eq_HR}), is very sensitive to absorption, and thus capable of separating X--ray type-1 (unobscured or unabsorbed) from X--ray type-2 (obscured or absorbed) AGN (Mainieri et al. 2002; Della Ceca et al. 2004; Perola et al. 2004; Caccianiga et al. 2004; Dwelly et al. 2005; Hasinger et al 2008). Full details on the methodology applied are given in Povi\'c et al. (2009). Following Fiore et al. (2003), all objects having X/O\,$>$\,0.1 and HR$(2-4.5$keV/$0.5-2$keV$)$\,$<$\,-\,0.35 have been classified as X--ray type-1 (unobscured), while objects with X/O\,$>$\,0.1 and HR$(2-4.5$keV/$0.5-2$keV$)$\,$>$\,-\,0.35 have been classified as X--ray type-2 (obscured) AGN. Figure~\ref{im_logXO_HRh2s_sxds_gws_morph} shows the X--ray type classification diagram including the host morphological information (compact, early- and late-type galaxies). A large fraction of objects (59.3\%) fall inside the region of X--ray type-1 AGN. The dashed--line box indicates the `highest probability' region for finding X--ray type-1 AGN (Della Ceca et al. 2004). 34.9\% of our sample (SXDS\,+\,GWS) fall in this box while only 11\% fall in the lower region of the diagram (i.e. typical of stars, normal galaxies and Compton-thick AGN). No evidence of any relationship between X--ray and morphological types was indicated since there was no clear separation of X--ray properties based on morphological types. Nevertheless, it can be seen that the majority (more than 60\%) of objects identified as compact are placed in the X--ray type-1 region.

\section{XLFs of different AGN populations}
\label{sec_LF}

\subsection{LF measurements}
\label{sec_LF_measure}
 
\indent \indent The LF is one of the most important tools for studying the 
evolution of extragalactic sources. It is defined as the number of objects per unit comoving volume 
per unit luminosity interval, i.e.
\begin{equation} \label{eq_phi}
{\Phi(L,z) = {\frac{\displaystyle d^2N}{\displaystyle dV dL}} (L,z)}, 
\end{equation}
where $N$ is the number of objects of luminosity $L$ found in the comoving volume $V$ at redshift $z$. In calculating the LF, we used the 1/$V_a$ method (Schmidt 1968) to measure the space density ($dN/dV$) as:
\begin{equation}\label{eq_dNdV}
{\frac{\displaystyle dN}{\displaystyle dV}(L,z) \approx \displaystyle\sum_{i=1}^N \frac{\displaystyle 1}{\displaystyle V_a(i)}}, 
\end{equation}
with $N$ objects having the luminosities and redshifts in the interval $\Delta L \Delta z$ around the centre of the bin ($L,z$). $V_a(i)$ is the survey available comoving volume within which an 
object $i$ with luminosity $L(i)$ could be detected and remained in the redshift bin $\Delta z$. The main advantage of the $dN/dV$ estimator (compared with simply dividing the number of objects found by the average search volume) is that it takes into account the fact that more luminous objects are detectable over larger volumes 
than less luminous objects. \\
\indent To derive the LF ($\Phi$) through the space density in the observed luminosity bin and surveyed solid angle ($\Omega$) we used the following equation: 
\begin{equation} \label{eq_phi}
{\Phi(logL,z) = {\frac{\displaystyle 4\pi}{\displaystyle \Omega}}{\frac{\displaystyle 1}{\displaystyle \Delta logL}} \displaystyle\sum_{i=1}^N \frac{\displaystyle 1}{\displaystyle V_a(i)}}. 
\end{equation}
A bin of logarithmic luminosity has been used in this work. The available comoving volume $V_a(i)$ has been measured through the comoving distance (e.g., Hogg 2000), for the concordance cosmology assumed in this work.\\
\indent The statistical uncertainty of $\Phi$ was estimated using the following expression (e.g. Marshall 1985; Boyle, Shanks \& Peterson 1988; Page \& Carrera 2000):
\begin{equation}\label{eq_Phi_err}
{\delta \Phi = {\frac{\displaystyle 1}{\displaystyle \Delta L}} \sqrt {\sum_{i=1}^N \Bigg(\frac{\displaystyle 1}{\displaystyle V_a(i)}\Bigg)^2}}. 
\end{equation}

\begin{figure*}[t]
\centering
\vspace*{1truecm}
\includegraphics[width=17.1cm,angle=0]{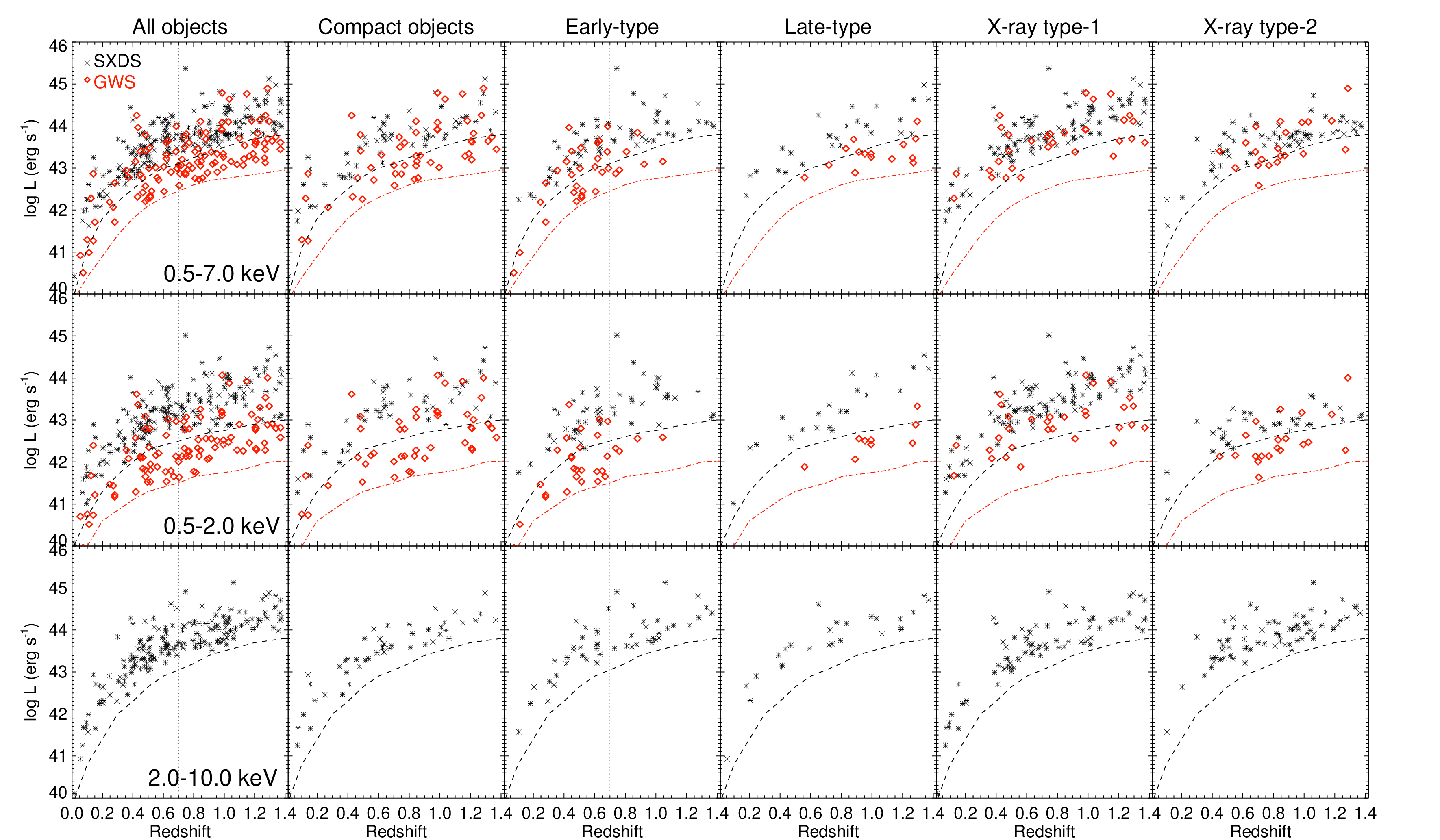}
\protect\caption[ ]{\textit{From top to bottom:} Relation between the X--ray luminosity in the 0.5\,-\,7.0\,keV, 0.5\,-\,2.0\,keV, and 2.0\,-\,10.0\,keV bands and redshift for a sample of X--ray emitters with optical counterparts in the SXDS (black stars) and GWS (red diamonds) fields. \textit{From left to right:} This relation is shown for the whole sample of objects, and for different morphological 
(compact, early- and late-type galaxies) and X--ray types (X--ray type-1 and type-2 sources). Vertical thin, dotted line defines the limit between two redshift intervals that we used to study the evolution in Sections~\ref{sec_LF_Xray_wholesample} and ~\ref{subsec_LF_Xray_Xrayty1}. Black dashed and red dash-dot-dashed lines present the luminosity limits of the SXDS and GWS samples, respectively. 
\label{im_Lz_xray}}
\end{figure*}

A Schechter (1976) function has been used to fit the LF:
\begin{equation}\label{eq_schechter_L}
{\Phi(L)dL = \Phi^* \Bigg({\frac{\displaystyle L}{\displaystyle L^*}}\Bigg)^{\alpha} exp {\Bigg({-\frac{\displaystyle L}{\displaystyle L^*}}\Bigg)} \frac{\displaystyle dL}{\displaystyle L^*}}, 
\end{equation}
where $\Phi(L)dL$ is the number of galaxies with luminosity between $L$ and $L$\,+\,$dL$ per Mpc$^3$. This approximation has been widely used for more than 30 year and provides a good fit for both local (e.g. Loveday et al. 1992; Blanton et al. 2001) and high redshift sources (e.g. Norberg et al. 2002; Faber et al. 2007; Ouchi et al. 2008; 
Dai et al. 2009). The Schechter function has three free 
parameters which must be determined empirically: $\alpha$, $\Phi^*$ and $L^*$. The quantity $\alpha$ defines the faint end slope of the LF, and indicates how quickly galaxy number counts evolve at low luminosities. It is usually negative, 
implying a large number of galaxies with low luminosities; if $\alpha$\,=\,$-$1 the LF is said to be flat. $L^*$ is the so-called characteristic Schechter 
luminosity, and $\Phi^*$ is the number of galaxies per Mpc$^3$ per luminosity at the characteristic luminosity. \\
\indent We computed the best fit Schechter parameters using a Marquardt-Levenberg 
algorithm (Levenberg 1944; Marquardt 1963) by means of the IDL LMFIT\footnote{http://idlastro.gsfc.nasa.gov/idl\_html\_help/LMFIT.html\#wp831462} function, performing up to 100 iterations. Finally, the best fit parameters and their 1$\sigma$ uncertainty estimates, assuming Poisson statistics, were obtained.\\

\subsection{LF of whole AGN sample}
\label{sec_LF_Xray_wholesample}

\indent \indent Our XLF analysis has been restricted to sources with z\,$\le$\,1.4 in order to have a sample with reliable redshifts and an optimal number of sources in the L$_X$-z plane (Figure~\ref{im_Lz_xray}). Figure~\ref{im_Lz_xray} shows this plane in the total, and soft X--rays for AGN in both the SXDS and GWS fields, and in the hard band for the AGN belonging to the SXDS field only (since hard range was not selected in the GWS field). In both the soft and total bands, the objects with lowest X--ray luminosities belong to the GWS field, since the GWS \textit{Chandra} data are deeper in comparison with the SXDS XMM-\textit{Newton} data (see Section~\ref{subsec_xray}). The diagrams show the X--ray luminosity as a function of redshift for the entire source sample and for different morphological and X--ray types (from left to right). Luminosity limits in all three bands for both SXDS and GWS samples are also presented. In all bands we can clearly see the rapid increase of luminosity at lower redshifts, and flatter luminosity distribution at higher redshifts. Also, we can see that the most luminous objects are the most distant. At 0.7\,$<$\,z$_2$\,$<$\,1.4, we found more compact sources in the total and soft bands than in the hard X--rays. Early-type objects dominate late-types at lower redshifts in the total and soft bands, but have similar populations in the hard band. 

\begin{figure*}[t]
\centering
\begin{minipage}[c]{.47\textwidth}
\includegraphics[width=8.0cm,angle=0]{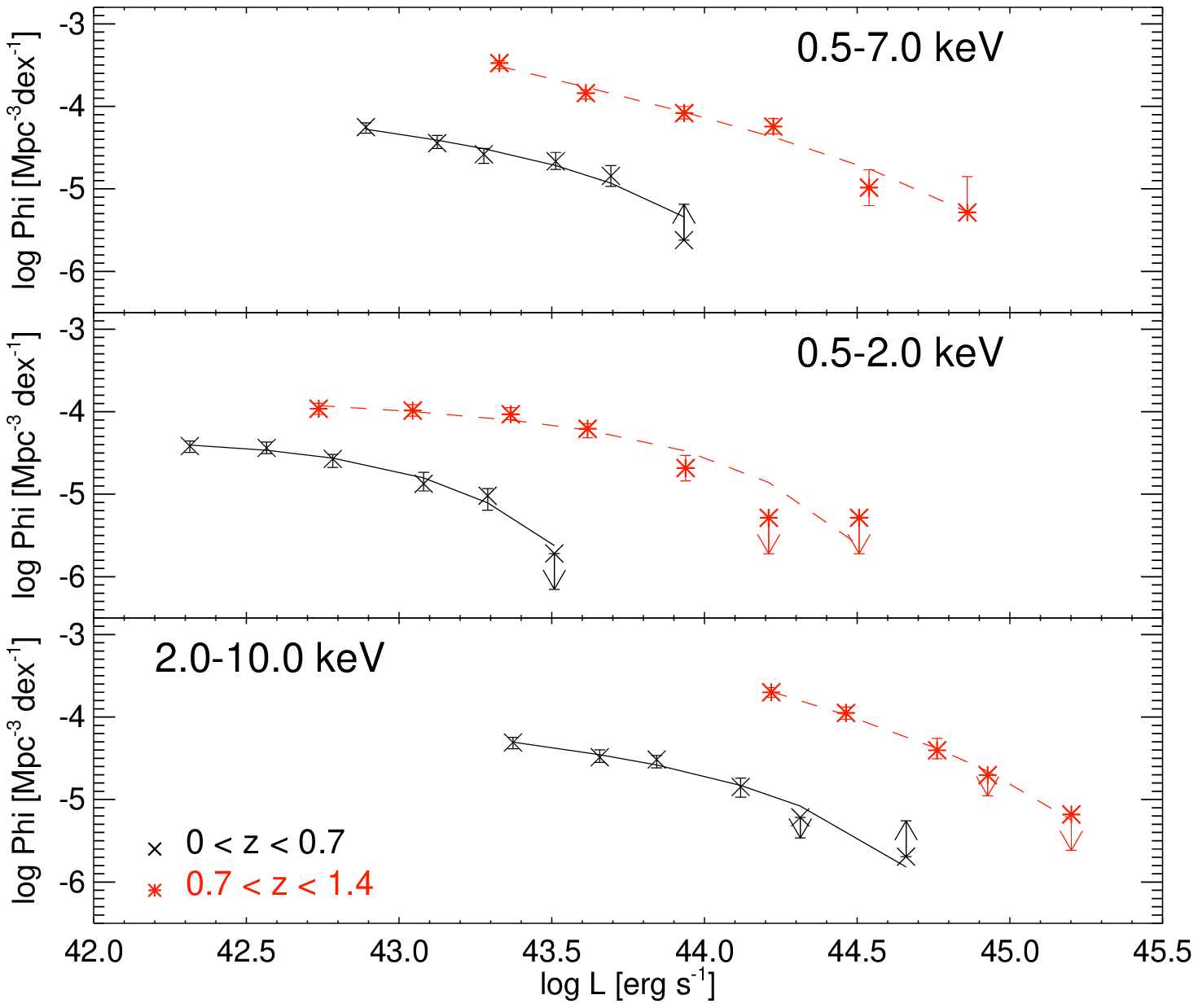}
\centering
\end{minipage}
\begin{minipage}[c]{.47\textwidth}
\includegraphics[width=8.0cm,angle=0]{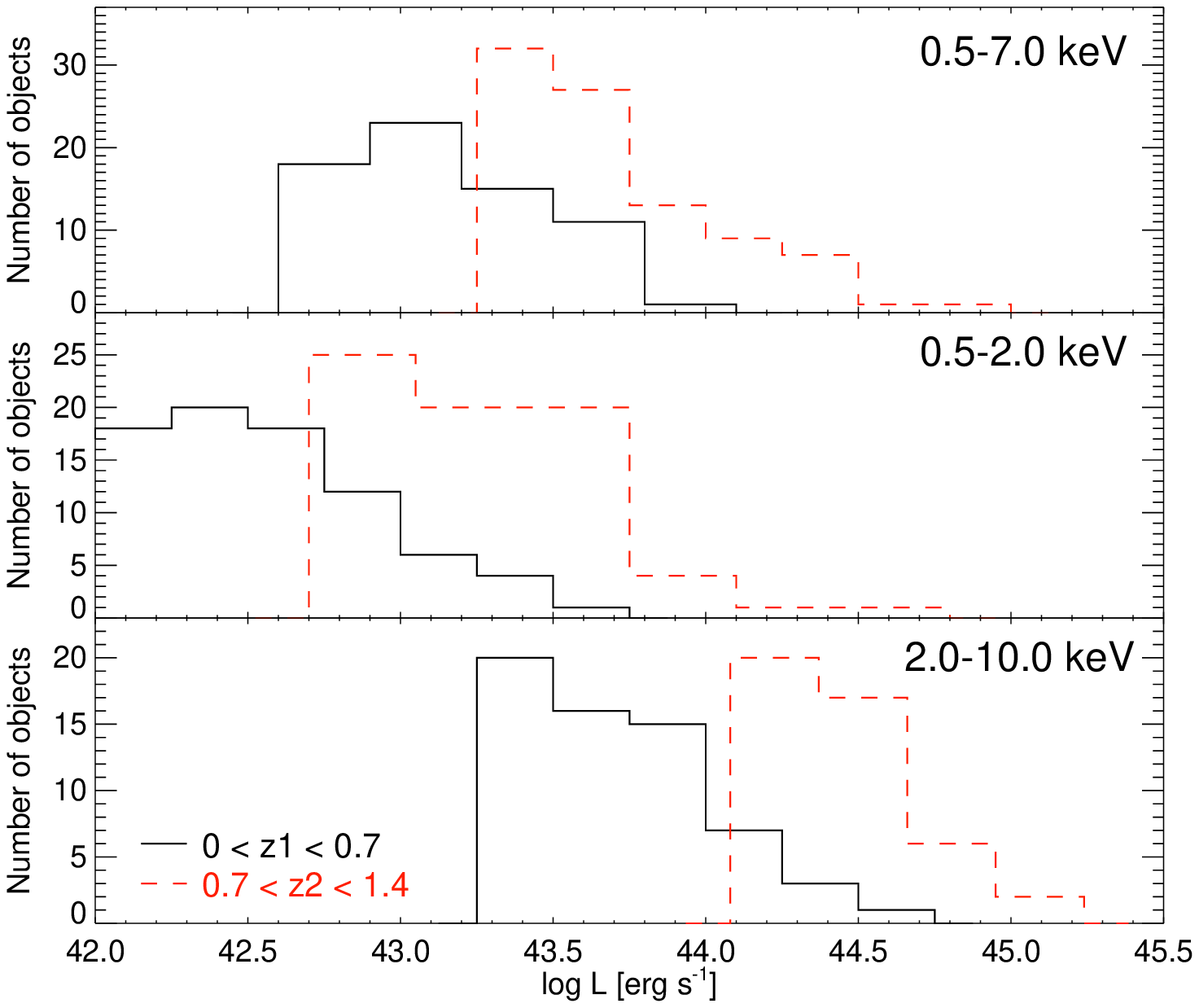}
\centering
\end{minipage}
\caption[ ]{\textit{(Left, from top to bottom)}: TXLF, SXLF, and HXLF in the z$_1$\,$\le$\,0.7 (black solid line, x symbols) and 0.7\,$\le$\,z$_2$\,$\le$\,1.4 (red dashed line, stars) intervals. The best-fit Schechter parameters are shown in Table~\ref{tab_schechter_X_totsample}. The arrows represent 1$\sigma$ errors in those luminosity bins where we have data only above or below the central luminosity. In all following figures, the arrows have the same significance. \textit{(Right, from top to bottom)}: Luminosity distribution in each redshift bin in the total, soft, and hard bands. 
\label{fig_lf_xrays_histos_allbands}}
\end{figure*}

\begin{table*}[]
\begin{center}
\caption{The best-fit Schechter parameters of the XLF for a whole sample of AGN (Figure~\ref{fig_lf_xrays_histos_allbands})
\label{tab_schechter_X_totsample}}
\small{
\begin{tabular}{c| c c c c c c c}    
\noalign{\smallskip}
\hline\hline
\noalign{\smallskip}
\textbf{Energy range}&\textbf{z$^a$}&\textbf{Num$^{b}$}&\textbf{log$\Phi^*$}&\textbf{$\alpha$}&\textbf{log$L^*$}&\\
\textbf{(keV)}&&&\textbf{(Mpc$^{-3}$)}&&\textbf{(erg\,s$^{-1}$)}\\
\noalign{\smallskip}
\hline\hline
\noalign{\smallskip}
\textbf{0.5\,-\,7.0}&0\,-\,0.7&127&-4.49\,$\pm$\,0.66&-1.35\,$\pm$\,0.57&43.69\,$\pm$\,0.51\\
&0.7\,-\,1.4&126&-4.76\,$\pm$\,0.80&-1.85\,$\pm$\,0.24&44.82\,$\pm$\,0.59\\
\noalign{\smallskip}
\hline
\noalign{\smallskip}
\textbf{0.5\,-\,2.0}&0\,-\,0.7&111&-4.31\,$\pm$\,0.31&-0.99\,$\pm$\,0.44&43.03\,$\pm$\,0.26\\
&0.7\,-\,1.4&99&-4.09\,$\pm$\,0.22&-1.15\,$\pm$\,0.20&43.98\,$\pm$\,0.18\\
\noalign{\smallskip}
\hline
\noalign{\smallskip}
\textbf{2.0\,-\,10.0}&0\,-\,0.7&97&-4.55\,$\pm$\,0.42&-1.35\,$\pm$\,0.38&44.25\,$\pm$\,0.30\\
&0.7\,-\,1.4&87&-4.32\,$\pm$\,1.16&-1.92\,$\pm$\,0.69&44.98\,$\pm$\,0.63\\
\noalign{\smallskip}
\hline
\hline
\end{tabular}
}
\begin{flushleft}
$^{a}$ Redshift range \\
$^{b}$ Number of objects \\
\end{flushleft}
\normalsize
\rm
\end{center}
\end{table*}

\indent XLFs derived in the 0.5\,-\,7.0 keV (TXLF), 0.5\,-\,2.0\,keV (SXLF), and 2.0\,-\,10.0\,keV (HXLF) bands are presented in Figure~\ref{fig_lf_xrays_histos_allbands} (left panel, from top to bottom, respectively). The LFs were analysed in two volumes with similar numbers of objects in all bands.
The first volume corresponds to the redshift range 0\,$<$\,z$_1$\,$\le$0.7 that represents about 1/4 of the total volume analysed ($\approx$\,0.009 Gpc$^3$ for a covered solid angle), while the second volume corresponds to the redshift range 0.7\,$<$\,z$_2$\,$\le$1.4 occupying the rest of the volume. The TXLF and SXLF were derived from both SXDS and GWS samples to flux limits of 9.0\,$\times$\,10$^{-15}$ and 
2.25\,$\times$\,10$^{-15}$\,erg cm$^{-2}$ s$^{-1}$ in each band, respectively. Since the X--ray GWS data are deeper than the SXDS data (see Section~\ref{subsec_xray}), the used limiting fluxes correspond to the completeness levels of the SXDS samples. The HXLF has been derived from the SXDS sample to a flux limit of 4.0\,$\times$\,10$^{-15}$\,erg\,cm$^{-2}$\,s$^{-1}$. The distributions of objects 
belonging to each observed luminosity bin in the total, soft, and hard bands are shown in Figure~\ref{fig_lf_xrays_histos_allbands} (right panel, from top to bottom, respectively). \\
\indent In order to estimate the error in the derivation of the LF due to cosmic variance when both the SXDS and GWS samples are analysed, the Cosmic Variance Calculator\footnote{http://casa.colorado.edu/$\sim$trenti/CosmicVariance.html} (Trenti \& Stiavelli 2008) was used. Using both the Press-Schech-ter (Press \& Schechter 1974) and the Sheth-Tormen (Sh-eth \& Tormen 1999) models, the error in the LF due to cosmic variance is smaller than 10\%.

\indent A good fit to the LF was obtained by using a Schechter function. The best fit Schechter parameters are given in Table~\ref{tab_schechter_X_totsample}. The shape of the LF remains similar 
in the three X--ray bands (in both redshift intervals). The faint-end slopes ($\alpha$) of the luminosity 
function does not show a significant difference with redshift. While 
the SXLF shows similar behaviour in z$_1$ and z$_2$ bins, it shows a shallower faint-end slope $\alpha$\,$\sim$\,-1.0, 
suggesting slower evolution than the other two bands. The plots of Figure~\ref{fig_lf_xrays_histos_allbands} taken 
together show that hard X--ray sources have higher luminosities than those 
selected in the 0.5-7.0\,keV band, a contrast even stronger in comparison with 
the soft X--ray objects. This difference is also found in the luminosity 
density with soft X--ray sources showing the lowest values.

\indent Analysing differences in evolution between z$_1$ and z$_2$ ranges 
we find:\\
\begin{itemize}
\item  [*] \textit{Luminosity Evolution}\\
Strong XLF was found in all bands as shown in Figure~\ref{fig_lf_xrays_histos_allbands}. 
Sources typically show lower luminosities in z$_1$ than in z$_2$ band. The Schechter fits in the three X--ray energy ranges lead to the 
same conclusion: $L^*$ shows evolution with redshift, having lower values in z$_1$ interval.
\item  [*] \textit{Density Evolution}\\
In general, there is no evidence for any significant AGN density evolution in any considered band. The 
sources become brighter as redshift increases, but 
given the uncertainties, their characteristic number density $\Phi^*$ 
associated to a characteristic luminosity ($L^*$) is consistent between the 
two analysed redshift intervals.  
\end{itemize}

\indent Strong AGN luminosity evolution was also found in previous studies (e.g. Miyaji et al. 2001; Ueda et al. 2003; La Franca et al. 2005; Hasinger et al. 2005; Silverman et al. 2008a; Ebrero et al. 2009; Yencho et al. 2009; Aird et al. 2010 and references 
therein). This evolution shows that a larger population of high-luminosity AGN is found at higher redshifts, while low-luminosity AGN are dominant at lower 
redshifts. This could be a result of: (i) selection effects (e.g., as redshift increases, first, the detectability of low-lumino-sity AGN decreases, and second, the observed volume increases, thus increasing the probability of detecting a lar-ger number of high-luminosity sources), (ii) true evolutionary effects, 
or (iii) a mixture of both. Selection effects will be tested in future, using a much larger sample of active galaxies, however the most luminous sources we observe at high redshifts we do not observe in the local universe.

\subsection{LF of compact, early- and late-type active galaxies}
\label{subsec_LF_Xray_morph}

\begin{figure}[t]
\centering
\vspace*{1truecm}
\includegraphics[width=8.3cm,angle=0]{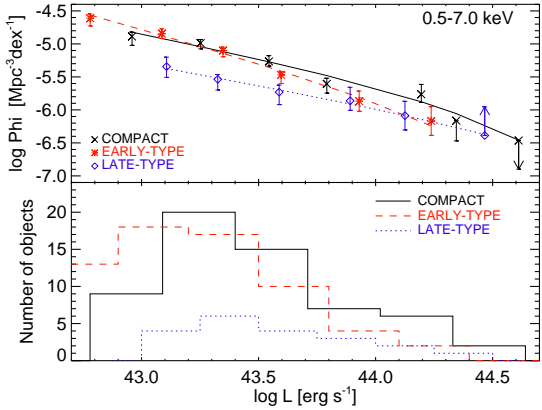}
\protect\caption[ ]{\textit{(Top)}: TXLF derived for different morphological types: compact (black crosses), 
early- (red stars) and late-type (blue diamonds) sources in z\,$\le$\,1.4 redshift interval. \textit{(Bottom)}: Distribution of objects in each luminosity 
bin of compact (black line), early- (red line) and late-type (blue line) objects. The best-fit Schechter parameters are shown 
in Table~\ref{tab_schechter_X_morph}. 
\label{im_fl_xray_morph}}
\end{figure}

\indent \indent We also studied the AGN XLF in relation to morphology. The top panel in Figure~\ref{im_fl_xray_morph} shows the XLF as a function of morphology, while the bottom panel shows the luminosity distribution for each morphological class. The LFs of compact, early- and late-type objects were derived over the full 0.5\,-\,7.0\,keV range. Best-fit Schechter parameters are given in Table~\ref{tab_schechter_X_morph}. Due to the small number of objects in each morphological sample the LFs were derived without dividing into different redshift bins. Therefore, the evolution rates were not studied as a function of morphology in the framework of this paper. No significant differences were found within the uncertainties between LFs of compact and early-type objects in the total band. 
They have similar steep faint-end slopes and similar $L^*$ and $\Phi^*$ best-fit parameters. The situation may reverse for late-type objects showing a shallower faint-end slope, suggesting slower evolution, and higher break luminosity. However, a larger sample of late-type objects is necessary to confirm the results obtained. \\

\begin{table*}[ht!]
\begin{center}
\caption{The best-fit Schechter parameters of the TXLF as a function of morphology at z\,$\le$\,1.4 (Figure~\ref{im_fl_xray_morph})
\label{tab_schechter_X_morph}}
\small{
\begin{tabular}{c| c c c c}    
\noalign{\smallskip}
\hline\hline
\noalign{\smallskip}
\textbf{Sample}&\textbf{Num$^{a}$}&\textbf{log$\Phi^*$}&\textbf{$\alpha$}&\textbf{log$L^*$}\\
&&\textbf{(Mpc$^{-3}$)}&&\textbf{(erg\,s$^{-1}$)}\\
\noalign{\smallskip}
\hline\hline
\noalign{\smallskip}
\textbf{Compact}&79&-5.98\,$\pm$\,0.84&-1.72\,$\pm$\,0.29&44.59\,$\pm$\,0.64\\
\noalign{\smallskip}
\noalign{\smallskip}
\textbf{Early-type}&79&-6.20\,$\pm$\,1.37&-1.98\,$\pm$\,0.31&44.46\,$\pm$\,0.97\\
\noalign{\smallskip}
\noalign{\smallskip}
\textbf{Late-type}&29&-6.63\,$\pm$\,3.38&-1.65\,$\pm$\,0.57&45.04\,$\pm$\,3.68\\
\noalign{\smallskip}
\hline
\hline
\end{tabular}
}
\begin{flushleft}
$^{a}$ Number of objects \\
\end{flushleft}
\normalsize
\rm
\end{center}
\end{table*}

\indent These results are, in general, consistent with those of Georgakakis et al. (2009) who studied the contribution of different 
morphological types to the TXLF at redshifts 0.5\,-\,1.3 for the Extended Groth Strip field (EGS). Both this work and Georgakakis et al. (2009) found that relative to late-type objects, early-type hosts make a major contribution to the TXLF over the entire luminosity range, while at higher luminosities compact objects become more dominant.

\subsection{LF of X--ray type-1 and type-2 AGN}
\label{subsec_LF_Xray_nucty}

\begin{figure*}[t]
\centering
\begin{minipage}[c]{.47\textwidth}
\includegraphics[width=8.0cm,angle=0]{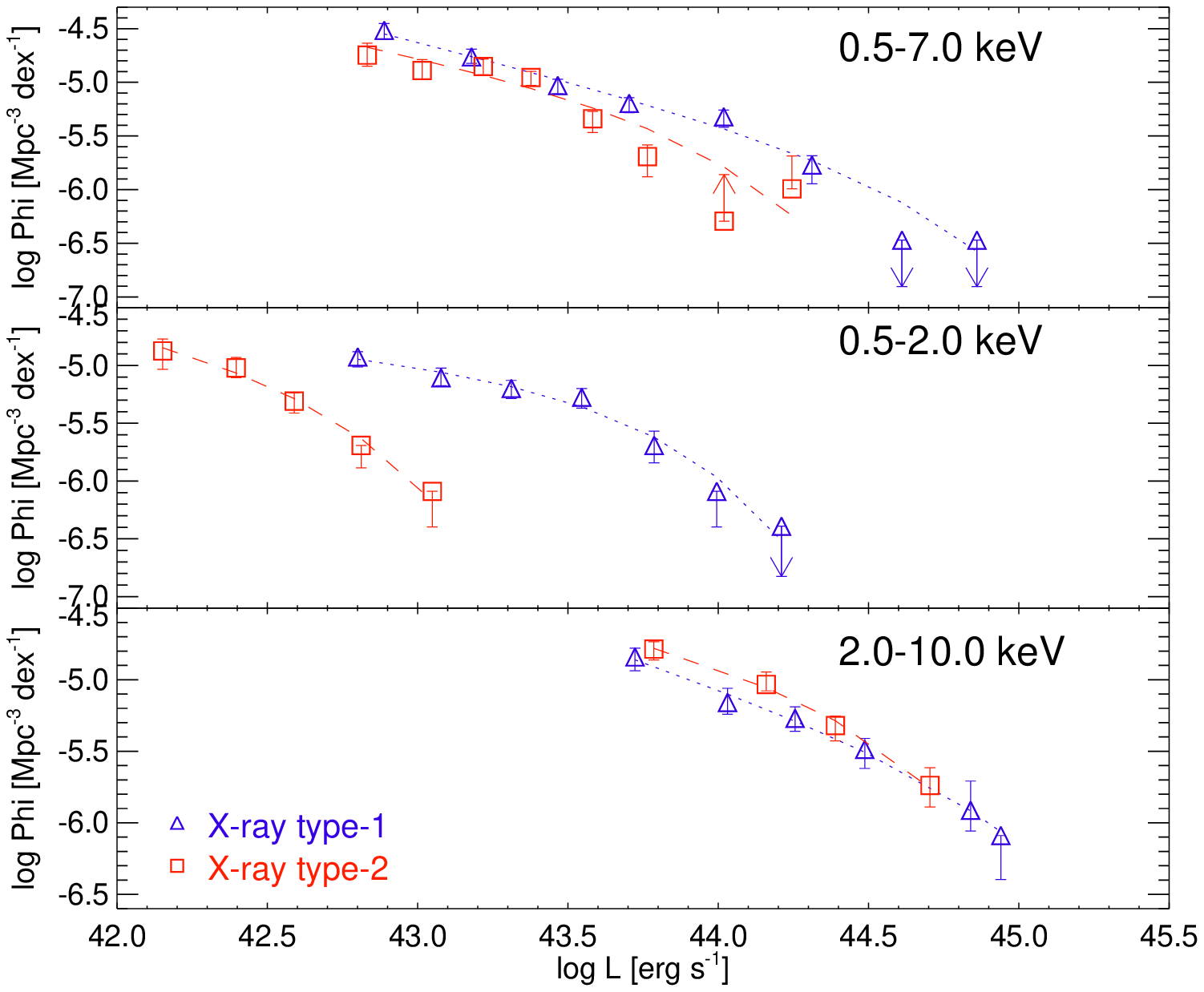}
\centering
\end{minipage}
\begin{minipage}[c]{.47\textwidth}
\includegraphics[width=8.0cm,angle=0]{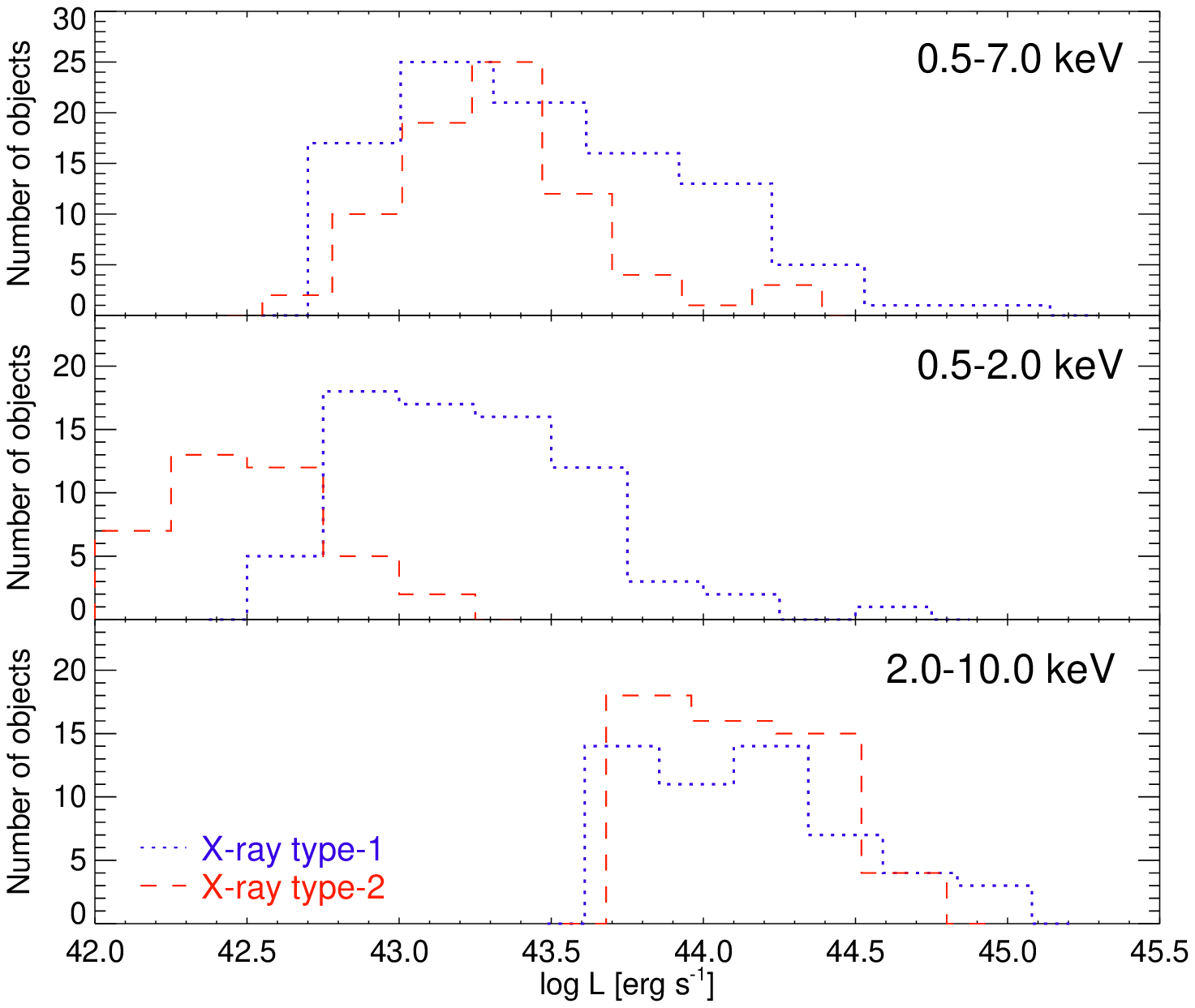}
\centering
\end{minipage}
\caption[ ]{\textit{(Left, from top to bottom)}: TXLF, SXLF, and HXLF of X--ray type-1 (blue triangles) and type-2 (red squares) objects at redshifts z\,$\le$\,1.4. The best-fit Schechter parameters are shown in Table~\ref{tab_schechter_X_Xty1ty2}. \textit{(Right, from top to bottom)}: Luminosity distribution in the total, soft, and hard bands of type-1 (blue doted line) and type-2 (red dashed line) sources. 
\label{im_fl_xray_Xty1ty2_tot2_soft_h2vh}}
\end{figure*}

\indent \indent We derive the XLF in relation to the X--ray obscuration. Figure~\ref{im_fl_xray_Xty1ty2_tot2_soft_h2vh} (left panel) shows the TXLF, SXLF, and HXLF (from top to bottom) of X--ray type-1 and type-2 sources. The numbers of sources in each luminosity bin of both types are shown for each energy range (right panel). 
Table~\ref{tab_schechter_X_Xty1ty2} summarises the derived best-fit Schechter parameters. Due to the small number of type-2 objects, the 
LFs of type-1 and type-2 sources were compared at all redshifts z\,$\le$\,1.4, without studying their evolution rates. Furthermore, in order to provide a good Schechter fit of the X--ray type-2 objects in the soft band, the value of the faint-end slope parameter was fixed to -1.5. Different values were tested, from -1.0 to -2.0 with a bin of 0.25, and the value 
of -1.5 was found to give the best fit. In the other two energy ranges the faint-end slope parameter stays free for both type-1 and type-2 objects.\\

\begin{table*}[ht!]
\begin{center}
\caption{The best-fit Schechter parameters of X--ray type-1 and type-2 sources at z\,$\le$\,1.4  (Figure~\ref{im_fl_xray_Xty1ty2_tot2_soft_h2vh})
\label{tab_schechter_X_Xty1ty2}}
\small{
\begin{tabular}{c| c| c c c c}    
\noalign{\smallskip}
\hline\hline
\noalign{\smallskip}
\textbf{En. range}&\textbf{Sample}&\textbf{Num$^{a}$}&\textbf{log$\Phi^*$}&\textbf{$\alpha$}&\textbf{log$L^*$}\\
\textbf{(keV)}&&&\textbf{(Mpc$^{-3}$)}&&\textbf{(erg\,s$^{-1}$)}\\
\noalign{\smallskip}
\hline\hline
\noalign{\smallskip}
\textbf{0.5\,-\,7.0}&\textbf{X--ray type-1}&120&-5.82\,$\pm$\,0.47&-1.71\,$\pm$\,0.15&44.69\,$\pm$\,0.37\\
\noalign{\smallskip}
\noalign{\smallskip}
&\textbf{X--ray type-2}&92&-5.29\,$\pm$\,0.47&-1.56\,$\pm$\,0.32&43.98\,$\pm$\,0.31\\
\noalign{\smallskip}
\hline
\noalign{\smallskip}
\textbf{0.5\,-\,2.0}&\textbf{X--ray type-1}&115&-5.13\,$\pm$\,0.31&-1.25\,$\pm$\,0.3&43.75\,$\pm$\,0.23\\
\noalign{\smallskip}
\noalign{\smallskip}
&\textbf{X--ray type-2}&50&-4.99\,$\pm$\,0.20&-1.50$^b$&42.68\,$\pm$\,0.15\\
\noalign{\smallskip}
\hline
\noalign{\smallskip}
\textbf{2.0\,-\,10.0}&\textbf{X--ray type-1}&82&-5.75\,$\pm$\,0.97&-1.70\,$\pm$\,0.40&45.01\,$\pm$\,0.72\\
\noalign{\smallskip}
\noalign{\smallskip}
&\textbf{X--ray type-2}&73&-5.05\,$\pm$\,0.65&-1.46\,$\pm$\,0.55&44.54\,$\pm$\,0.45\\
\noalign{\smallskip}
\hline
\hline
\end{tabular}
}
\begin{flushleft}
$^{a}$ Number of objects \\
$^{b}$ Fixed value of $\alpha$ in the Schechter fit \\
\end{flushleft}
\normalsize
\rm
\end{center}
\end{table*}

\indent The luminosity distributions of X--ray type-1 and type-2 objects appear to be similar at both total and hard bands, but significantly different in the soft band, as expected since soft X--rays are especially sensitive to absorption by the central torus. \\
\indent In general our results suggest that type-1 sources are more numerous at higher luminosities, and type-2 objects dominate the AGN 
population at lower luminosities. This is known as the `Steffen effect' (Steffen et al. 2003), and was observed in many later works 
(e.g., Ueda et al. 2003; Barger et al. 2005; Hasinger et al. 2008). Taking into account uncertainties, the faint-end slopes of X--ray 
type-1 and type-2 sources in the 0.5\,-\,7.0\,keV and 2.0\,-\,10.0\,keV bands appear to be similar at the observed redshifts. Moreover, observing the best Schechter-fit parameters, in general type-1 sources show higher characteristic luminosities ($L^*$) then the type-2 objects. In contrast, the values of the number density $\Phi^*$ at characteristic luminosity do not show significant differences between type-1 and type-2 sources in any of the three tested X--ray bands.

\subsection{Evolution of X--ray type-1 AGN}
\label{subsec_LF_Xray_Xrayty1}

\begin{figure}[t]
\centering
\vspace*{1truecm}
\includegraphics[width=8.3cm,angle=0]{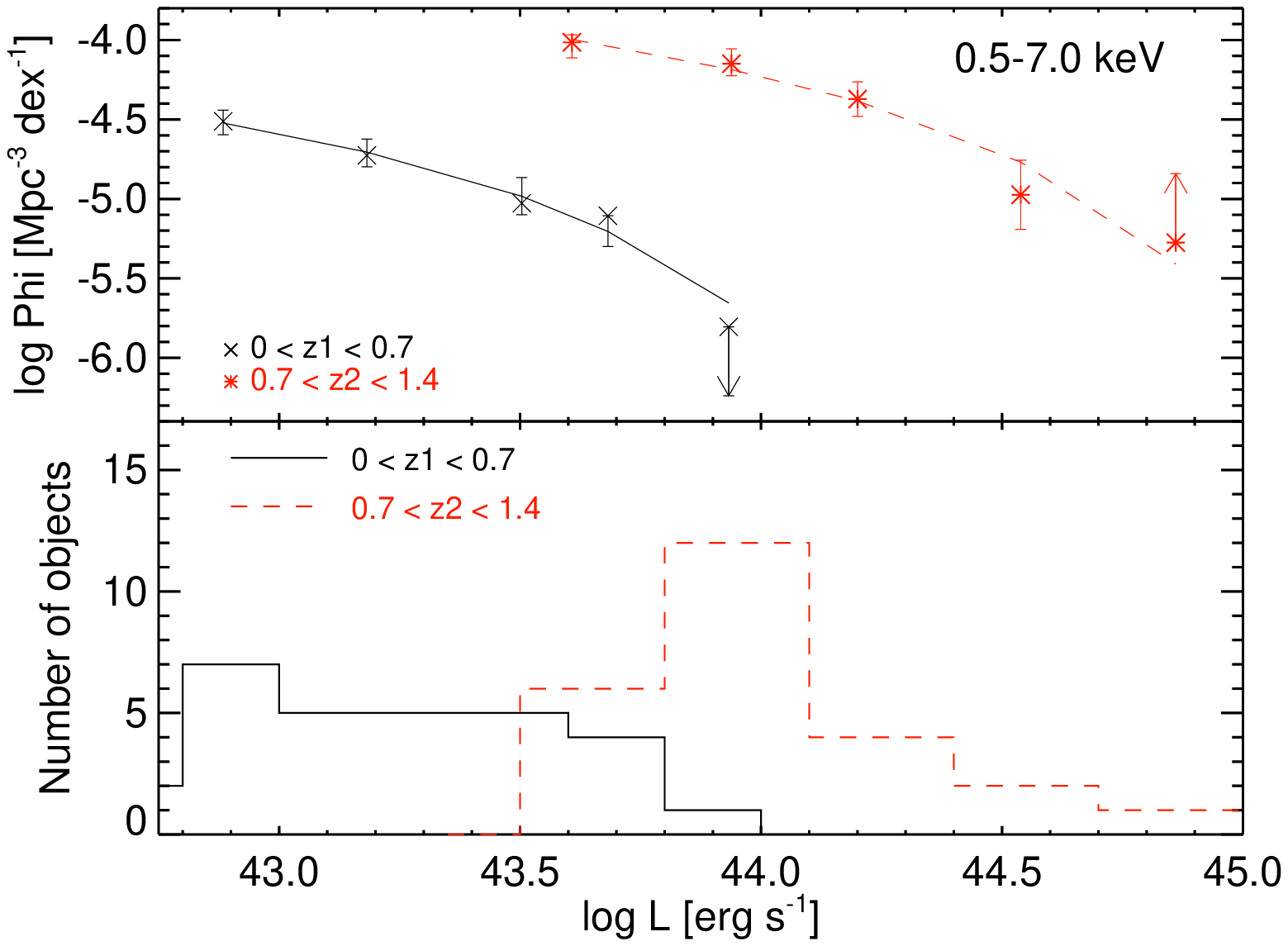}
\protect\caption[ ]{\textit{(Top)}: TXLF of X--ray type-1 sources derived in two redshift intervals: 
z$_1$\,$\le$\,0.7 (black crosses) and 0.7\,$<$\,z$_2$\,$\le$\,1.4 (red stars). \textit{(Bottom)}: Distribution of objects in each luminosity bin of X--ray type-1 
sources in the z$_1$\,$\le$\,0.7 (solid black line) and 0.7\,$<$\,z$_2$\,$\le$\,1.4 (dashed red line) redshift intervals. The best-fit Schechter parameters 
are shown in Table~\ref{tab_schechter_tot_Xty1_z1z2}. 
\label{im_fl_xray_Xty1_z1z2}}
\end{figure}

\indent \indent We studied the evolution of X--ray type-1 AGN. Figure~\ref{im_fl_xray_Xty1_z1z2} shows their TXLF (top panel), as well as the number of sources in each luminosity bin (bottom panel). The evolution rate was studied again in z$_1$ and z$_2$ intervals. 
Table~\ref{tab_schechter_tot_Xty1_z1z2} summarises the best-fit Schechter parameters.\\
\indent Similarly to the entire AGN  sample, evidence for a clear luminosity evolution was found with the most luminous objects residing at higher redshifts. 
Moreover, we observe a weak evolution in the number densities ($\Phi^*$), with high-redshift X--ray type-1 sources having higher $\Phi^*$ values.

\begin{table*}[ht!]
\begin{center}
\caption{The best-fit Schechter parameters of TXLF derived for X--ray type-1 sources (Figure~\ref{im_fl_xray_Xty1_z1z2})
\label{tab_schechter_tot_Xty1_z1z2}}
\small{
\begin{tabular}{c c c c c}    
\noalign{\smallskip}
\hline\hline
\noalign{\smallskip}
\textbf{z$^a$}&\textbf{Num$^{b}$}&\textbf{log$\Phi^*$}&\textbf{$\alpha$}&\textbf{log$L^*$}\\
&&\textbf{(Mpc$^{-3}$)}&&\textbf{(erg\,s$^{-1}$)}\\
\noalign{\smallskip}
\hline\hline
\noalign{\smallskip}
0\,-\,0.7&60&-4.74\,$\pm$\,0.76&-1.37\,$\pm$\,0.67&43.66\,$\pm$\,0.55\\
0.7\,-\,1.4&60&-4.28\,$\pm$\,0.53&-1.38\,$\pm$\,0.45&44.50\,$\pm$\,0.37\\
\noalign{\smallskip}
\hline
\hline
\end{tabular}
}
\begin{flushleft}
$^{a}$ Redshift range \\
$^{b}$ Number of objects \\
\end{flushleft}
\normalsize
\rm
\end{center}
\end{table*}

\section{Summary}
\label{sec_summary}
  
\indent \indent We present the XLF for X--ray detected AGN in the SXDS and GWS fields. We studied the entire sample
of active galaxies as well as different morphological and X--ray subsamples. In the future we will use a much larger sample of X--ray detected AGN from the COSMOS field to derive the LF in relation to the morphology of host galaxies and to study the evolution of different triggering mechanisms (work in progress).

\indent LFs of our entire AGN sample were derived in total, soft, and hard X--ray bands. We studied objects with z $\le$\,1.4 (where our photometric redshifts are most reliable relative to available spectroscopic redshifts) inferring source evolution in two redshift intervals (0\,$<$\,z$_1$\,$\le$\,0.7 and 0.7\,$<$ z$_2$ $\le$1.4) where each redshift band has a similar number of sources. The data were fitted with a Schechter function wh-ich was found to be a good approximation in all bands. \\
\indent The following results were obtained for the XLF derived for the entire sample:
\begin{itemize}
\item [$\star$] The LF shape does not change significantly between the two z intervals for the X--ray bands considered.  
\item [$\star$] Evidence for strong luminosity evolution was found in all X--ray bands, with higher density of more luminous objects at higher redshifts. 
\item [$\star$] Taking uncertainties into account, no significant signs of number density evolution was found in any of the tested bands. Sources become brighter as redshift increases but their characteristic number density ($\Phi^*$) that corresponds to characteristic luminosity $L^*$ appears to remain very much the same in both redshift intervals. 
\end{itemize}

\indent We studied the TXLF as a function of morphology for sources with z\,$\le$\,1.4. Three morphological types were tes-ted: compact, early-, and late-type with the following results:
\begin{itemize}
\item [$\star$] Compared to late-type objects, early-type hosts make a major contribution to the TXLF over the entire luminosity range, while at higher luminosities compact objects become more important. 
\item [$\star$] No significant differences \\were found between LFs of compact and early-type objects in the total X--ray band.
\end{itemize}

\indent The TXLF, SXLF, and HXLF of X--ray type-1 and type-2 sources at z\,$\le$\,1.4 were analysed, giving the following results:
\begin{itemize}
\item [$\star$] The SXLF of type-1 and type-2 sources are shown to be significantly different due to effects of absorption of soft X--rays by the AGN torus. In the total and hard bands the XLFs remain similar.  
\item [$\star$] We observed the `Steffen effect' in this work: type-1 sources are more numerous at higher luminosities, while type-2 objects are more common at lower luminosities. 
\item [$\star$] Type-1 sources show higher characteristic luminosities ($L^*$) than type-2 objects in the soft and total bands, whi-le at hard energies including uncertainties the properties stay similar.
\item [$\star$] Number densities $\Phi^*$ at characteristic luminosities do not show significant differences between X--ray type-1 and type-2 objects in any of three bands.
\end{itemize}

\indent Finally, the evolutionary rates of X--ray type-1 AGN we-re studied in z$_1$ and z$_2$ intervals. Deriving the LFs in each range, we obtained:
\begin{itemize}
\item [$\star$] The LFs show similar shape.
\item [$\star$] A clear evolution with luminosity was obtained, with more luminous AGN and higher luminosity densities at higher redshifts. We also observed a weak evolution in number densities, with higher redshift sources showing higher $\Phi^*$ values.
\end{itemize}

\acknowledgements
We thank Marc Huertas-Company for a number of valuable comments that improved the morphological classification of analysed sources. We also thank Jack Sulentic and the anonymous referee for all corrections and suggestions. We appreciate all comments from the XMM-\textit{Newton} team during our SXDS X--ray data reduction, and we also thank the SXDS team for making their data available to the astronomical community. This research has been supported by the Spanish Ministry of Economy and Competitiveness (MINECO) under the grant AYA2011-29517-C03-01. MP, IM, and JM acknowledge Junta de Andaluc\'ia and MINECO through projects PO8-TIC-03531 and AYA2010-15169. We acknowledge support from the Faculty of the European Space Astronomy Centre (ESAC). JIGS acknowledge financial support from the MINECO under project AYA2008-06311-C02-02 and AYA2011-29517-C03-02. JG acknowledges support from the MINECO through AYA2009-10368 project. The CEFCA is funded by the Fondo de Inversiones de Teruel, supported by both the Government of Spain (50\%) and the regional Government of Arag\'on (50\%). This work has been partially funded by the Spanish Ministerio de Ciencia e Innovaci-\'on through the PNAYA, under grants AYA2006-14056 and th-rough the ICTS 2009-14. This research has made use of software provided by the XMM-\textit{Newton} Science Operations Centre and Chandra X--ray Center (CXC) in the application packages SAS and CIAO, respectively. \texttt{IRAF} is distributed by the National Optical Astronomy Observatories, whi-ch are operated by the Association of Universities for Research in Astronomy, Inc., under cooperative agreement with the National Science Foundation. Virtual Observatory Tool for OPerations on Catalogues And Tables (TOPCAT) was used in this work.\\




\end{document}